# Speed-sensorless state feedback control of induction machines with LC filter

Julian Kullick[†] and Christoph M. Hackl[‡]

***Abstract*—A speed-sensorless state feedback controller for induction machines (IMs) with LC filter is proposed. The estimation of speed and remaining states is based on a speed-adaptive observer, requiring only the measurement of the filter input currents. The motor currents are controlled by a state-feedback controller with proportional and integral control action to achieve fast and asymptotic set point tracking. Observer *and* controller gains are calculated offline using linear quadratic regulator (LQR) theory and updated online (gain-scheduling), in order to guarantee stability and improve control performance in the whole operation range. The proposed control scheme is validated by simulation and experimental results including several zero-crossings of the mechanical speed. It is shown that the overall control system performs well under various load- and speed conditions; while its tuning remains easy making it attractive also for industrial application.**

## I. Introduction

In medium-voltage variable-speed drive applications with long power cables (e.g. large pumping systems), an inverter output (load) LC filter is often employed as to (i) decrease voltage deflection at the motor terminals, due to impedance imbalance between the cable and the motor, and to (ii) reduce steep voltage slopes which might damage the motor insulation and bearings due to high capacitive discharge [1]–[3].

However, the additional hardware comes at cost of electric coupling between the filter and motor currents and voltages, which in turn complicates the design of the overall control system. Therefore, in most industrial applications, the robust, easy-to-implement and speed-sensorless V/Hz-feedforward control strategy (scalar control) is used. However, it is less efficient and dynamic than vector control methods (e.g. field-oriented control (FOC)), in particular for variable-speed applications. Moreover, the induced torque ripples due to load changes and high inrush currents during startup in V/Hz-control put additional stress on the motor. Vector control could reduce mechanical and thermal stress on the machine components and, thereby, increase motor lifetime and efficiency. Moreover, incorporating sensorless control could make this control scheme appealing for applications, where the motor is located in hostile or rather inaccessible environments; such as electric submersible pump (ESP) systems [4].

Only few publications deal with sensor-based or sensorless vector control methods of IMs with LC filter. Kojima et al. first proposed a vector control scheme, using a cascade of deadbeat controllers for the filter input currents and output voltages and proportional-integral (PI) controllers for the motor currents, flux linkage and speed [5]. However, measurements of the rotational speed and all system states are presumed. In [6], Salomäki et al. present a speed-sensorless approach using a full-order adaptive observer with a modified adaption law, which solely requires measurements of the filter input currents. Their work is based on the approach of Hinkkanen and Luomi [7] and was later patented [8]. However, the employed control system is again based on cascaded PI-controllers, which require individual and non-intuitive tuning. Mukherjee and Poddar use an indirect method of speed estimation for their PI-controller based vector control scheme [9]. However, they presume measurements of the stator voltages and currents, which limits the approach to applications, where the motor terminals are accessible. Other approaches such as model predictive control (MPC) [10], direct torque control (DTC) [11] and feedback linearization [12] were also proposed.

In this paper, a state-feedback controller in combination with a speed-adaptive observer is applied to the considered ESP system for speed-sensorless control of an IM with LC filter. The main advantage of this approach is its simple implementation and easy tuning, making it particularly appealing for industrial applications. Moreover, stability of the controller and observer is guaruanteed by the linear quadratic regulator (LQR) design of the feedback gains in combination with gain scheduling to adapt to the actual system state (speed and load). The benefits of a simple and straight-forward observer and controller design (to be shown later) make up for its slightly higher computational burden on the real-time system. Especially for medium-voltage applications, the sampling time of the VSI is often low, which ensures enough computational time to complete the computation of the control action. The proposed control system is verified simulatively and experimentally on a down-scaled electrical drive system with LC filter, including several speed zero-crossings and drastic load variations.

The main contributions of this work are: (i) application of the LQR design method (with gain-scheduling) for the gain selection of the speed-adaptive observer, (ii) derivation of a state-feedback control scheme with integral action and prefilter

[†]J. Kullick is with the research group "Control of renewable energy systems" (CRES) at the Munich School of Engineering (MSE), Technical University of Munich (TUM), Germany.

[‡]C. M. Hackl is with the Department of Electrical Engineering and Information Technology at the Munich University of Applied Sciences (MUAS), Germany.

Funding from the Bavarian State Ministry of Education, Science and the Arts in the frame of the project Geothermie-Allianz Bayern is gratefully acknowledged.

for the given drive system (again, using LQR design with gain-scheduling) and (iii) practical recommendations on implementation matters, such as (a) discretization of the continuous-time control scheme and (b) recommendations for a simple but effective selection of the LQR weighting matrices.

In anticipation of the following sections, an overview of the control system, is given in Fig. 1, showing the physical system [- - -], the control system [- - -] and the observer system [- - -]. The corresponding equations of the observer system will be derived in Sec. III, while the controller equations will be derived in Sec. IV.

## II. SYSTEM DESCRIPTION AND STATE SPACE MODEL

The complete electrical drive system is depicted in the upper part of Fig. 1 (physical system). However, only the following components are considered in the drive model: (i) Voltage source inverter (VSI), (ii) inverter output filter (LC filter) and (iii) three-phase squirrel-cage IM.

The VSI produces modulated phase voltages $\boldsymbol{u}_{\mathrm{f}}^{abc}$ at its ouput terminals (according to the implemented modulation scheme). The phase currents $\boldsymbol{i}_{\mathrm{f}}^{abc}$ flow from the inverter output to the LC filter input. The LC filter output is in turn connected to the machine terminals. The (filtered) stator voltages $\boldsymbol{u}_{\mathrm{s}}^{abc}$ drive the stator currents $\boldsymbol{i}_{\mathrm{s}}^{abc}$. Due to magnetic coupling between stator and rotor of the machine, the induced voltages in the rotor windings (cage) produce currents resulting in the rotor flux linkage $\boldsymbol{\psi}_{\mathrm{r}}^{abc}$. The IM rotates at mechanical angular velocity $\omega_{\mathrm{m}}$, which is proportional to the electrical angular velocity $\omega_{\mathrm{r}} = n_{\mathrm{p}}\omega_{\mathrm{m}}$ with number $n_{\mathrm{p}}$ of IM pole pairs. The produced electromagnetic torque is denoted by $m_{\mathrm{e}}$ (load torque $m_{\mathrm{l}}$ and friction torque $m_{\mathrm{f}}$ act against $m_{\mathrm{e}}$).

In Fig. 1, all electrical quantities of the real system (upper part) are given in three-phase $abc$-coordinates. However, the modeling and control (lower part) will be conducted using space vector notation in the rotating $dq$-reference frame which is aligned with the rotor flux linkage (details are omitted). Hence, the $dq$-reference frame rotates at synchronous speed $\omega_k$ and is displaced by the angle $\phi_k$ from the stationary $\alpha\beta$-reference frame. Note that the time-dependency of the used quantities is not explicitly stated as to improve readability.

The following assumptions are imposed on the system.

**Assumption 1** (Magnetic saturation)**.** *Magnetic saturation is negligible and flux linkages depend linearly on the currents.*

**Assumption 2** (Quasi-constant speed)**.** *The mechanical system is significantly slower than the electrical system and, hence, $\omega_{\mathrm{m}}$ can be considered a slowly time-varying parameter.*

**Assumption 3** (Quasi-constant load)**.** *The load torque $m_{\mathrm{l}}$ is a slowly varying disturbance and, hence, the synchronous speed $\omega_k$ becomes a slowly time-varying parameter, too.*

**Assumption 4** (Measured quantities)**.** *Only dc link voltage $u_{\mathrm{dc}}$ and filter (input) currents $\boldsymbol{i}_{\mathrm{f}}^{abc}$ are measured and available for feedback.*

Based on Assumptions 1, 2, 3 & 4, the dynamics of IM and LC filter can be derived in the $dq$-reference frame as (see [4])

$$\left.\begin{aligned} \tfrac{\mathrm{d}}{\mathrm{d}t}\boldsymbol{x}^{dq} &= \boldsymbol{A}_{\mathrm{x}\to\mathrm{x}}(\omega_{\mathrm{r}},\omega_k)\boldsymbol{x}^{dq} + \boldsymbol{A}_{\mathrm{u}\to\mathrm{x}}\boldsymbol{u}_{\mathrm{f}}^{dq}, \\ \boldsymbol{y}^{dq} &= \boldsymbol{C}_{\mathrm{x}}\boldsymbol{x}^{dq} (= \boldsymbol{i}_{\mathrm{f}}^{dq}) \end{aligned}\right\} \tag{1}$$

where $\boldsymbol{x}^{dq} := (\boldsymbol{i}_{\mathrm{f}}^{dq\top}, \boldsymbol{u}_{\mathrm{s}}^{dq\top}, \boldsymbol{i}_{\mathrm{s}}^{dq\top}, \boldsymbol{\psi}_{\mathrm{r}}^{dq\top})^\top \in \mathbb{R}^8$ and $\boldsymbol{y}^{dq}$ are the system state vector and output, resp., and

$$\left.\begin{aligned} \boldsymbol{A}_{\mathrm{x}\to\mathrm{x}}(\omega_{\mathrm{r}},\omega_k) &:= \begin{bmatrix} -\frac{1}{T_{\mathrm{f}}}\boldsymbol{I}_2 & -\frac{1}{L_{\mathrm{f}}}\boldsymbol{I}_2 & \boldsymbol{0}_{2\times2} & \boldsymbol{0}_{2\times2} \\ \frac{1}{C_{\mathrm{f}}}\boldsymbol{I}_2 & \boldsymbol{0}_{2\times2} & -\frac{1}{C_{\mathrm{f}}}\boldsymbol{I}_2 & \boldsymbol{0}_{2\times2} \\ \boldsymbol{0}_{2\times2} & \frac{1}{\sigma L_{\mathrm{s}}}\boldsymbol{I}_2 & -\frac{1}{\widetilde{T}_{\mathrm{s}}}\boldsymbol{I}_2 & -\frac{1}{\widetilde{L}_m}(\omega_{\mathrm{r}}\boldsymbol{J} - \frac{1}{T_{\mathrm{r}}}\boldsymbol{I}_2) \\ \boldsymbol{0}_{2\times2} & \boldsymbol{0}_{2\times2} & \frac{L_m}{T_{\mathrm{r}}}\boldsymbol{I}_2 & \omega_{\mathrm{r}}\boldsymbol{J} - \frac{1}{T_{\mathrm{r}}}\boldsymbol{I}_2 \end{bmatrix} - \omega_k\boldsymbol{J}_8, \\ \boldsymbol{A}_{\mathrm{u}\to\mathrm{x}} &:= \begin{bmatrix} \frac{1}{L_{\mathrm{f}}}\boldsymbol{I}_2 & \boldsymbol{0}_{2\times2} & \boldsymbol{0}_{2\times2} & \boldsymbol{0}_{2\times2} \end{bmatrix}^\top, \\ \boldsymbol{C}_{\mathrm{x}} &:= \begin{bmatrix} \boldsymbol{I}_2 & \boldsymbol{0}_{2\times2} & \boldsymbol{0}_{2\times2} & \boldsymbol{0}_{2\times2} \end{bmatrix}, \end{aligned}\right\} \tag{2}$$

denote the system matrix ($\boldsymbol{A}_{\mathrm{x}\to\mathrm{x}} \in \mathbb{R}^{8\times8}$), input matrix ($\boldsymbol{A}_{\mathrm{u}\to\mathrm{x}} \in \mathbb{R}^{2\times8}$), output matrix ($\boldsymbol{C}_{\mathrm{x}} \in \mathbb{R}^{2\times8}$), resp., and $\boldsymbol{J}_8 \in \mathbb{R}^{8\times8}$ is a block matrix with $\boldsymbol{J} := \left[\begin{smallmatrix} 0 & -1 \\ 1 & 0 \end{smallmatrix}\right]$ matrices as block diagonal elements. Moreover, $T_{\mathrm{f}} := \frac{L_{\mathrm{f}}}{R_{\mathrm{f}_1}}, \widetilde{T}_{\mathrm{s}} := \frac{\sigma L_{\mathrm{s}}}{R_{\mathrm{s}}}$ and $T_{\mathrm{r}} := \frac{L_{\mathrm{r}}}{R_{\mathrm{r}}}$ are the filter, stator and rotor time constants; $L_{\mathrm{f}}, C_{\mathrm{f}}$ and $R_{\mathrm{f}}$ are the filter inductance, capacitance and series resistance; $L_{\mathrm{s}} := L_m + L_{\mathrm{s}\sigma}$, $L_{\mathrm{r}} := L_m + L_{\mathrm{r}\sigma}$, $L_m$, $L_{\mathrm{s}\sigma}$ and $L_{\mathrm{r}\sigma}$ are the stator and rotor self inductances, the main inductance and the stator and rotor leakage inductances; $\sigma := 1 - \frac{L_m^2}{L_{\mathrm{s}}L_{\mathrm{r}}}$ is the leakage coefficient; $R_{\mathrm{s}}$ and $R_{\mathrm{r}}$ are the stator and rotor resistances and $\widetilde{L}_m := \frac{\sigma L_{\mathrm{s}} L_{\mathrm{r}}}{L_m}$ is an auxiliary inductance term (for details, see [4]).

## III. OBSERVER SYSTEM

The observer system is depicted at the bottom right of Fig. 1 and combines various subcomponents that each contribute to reproducing the actual system state. The observer outputs (estimated quantities) are passed on to the control system, while the controller output and the filter current measurements serve as inputs.

### A. Inverter approximation

The VSI produces the modulated output voltage $\boldsymbol{u}_{\mathrm{f}}^{dq}$ according to the reference vector $\boldsymbol{u}_{\mathrm{f}}^{dq\star}$ by pulse width modulation (PWM). This output voltage generation comes with a time delay $t_{\mathrm{dt}}$, which depends on the employed modulation scheme and is inversely proportional to the switching frequency of the inverter. For medium voltage drive applications, typically low switching frequencies are used, which increases the time delay and thus necessitates its consideration in the model. The time delay is typically approximated by a first-order lag system [13, Chap. 14], yielding the simplified inverter model

$$\tfrac{\mathrm{d}}{\mathrm{d}t}\hat{\boldsymbol{u}}_{\mathrm{f}}^{dq} = \underbrace{-\left(\tfrac{1}{t_{\mathrm{dt}}}\boldsymbol{I}_2 + \omega_k\boldsymbol{J}\right)}_{=:\boldsymbol{A}_{\mathrm{u}\to\mathrm{u}}(\omega_k)}\hat{\boldsymbol{u}}_{\mathrm{f}}^{dq} + \underbrace{\tfrac{1}{t_{\mathrm{dt}}}\boldsymbol{I}_2}_{=:\boldsymbol{B}}\boldsymbol{u}_{\mathrm{f}}^{dq\star} \tag{3}$$

with inverter system state vector $\boldsymbol{x}_{\mathrm{u}}^{dq} \in \mathbb{R}^2$, input vector $\boldsymbol{u}^{dq} \in \mathbb{R}^2$, (electrical) speed dependent system matrix $\boldsymbol{A}_{\mathrm{u}\to\mathrm{u}}(\omega_k) \in \mathbb{R}^{2\times2}$, input matrix $\boldsymbol{B} \in \mathbb{R}^{2\times2}$, unity matrix $\boldsymbol{I}_2 := \left[\begin{smallmatrix} 1 & 0 \\ 0 & 1 \end{smallmatrix}\right]$ and rotation matrix $\boldsymbol{J} := \left[\begin{smallmatrix} 0 & -1 \\ 1 & 0 \end{smallmatrix}\right]$.

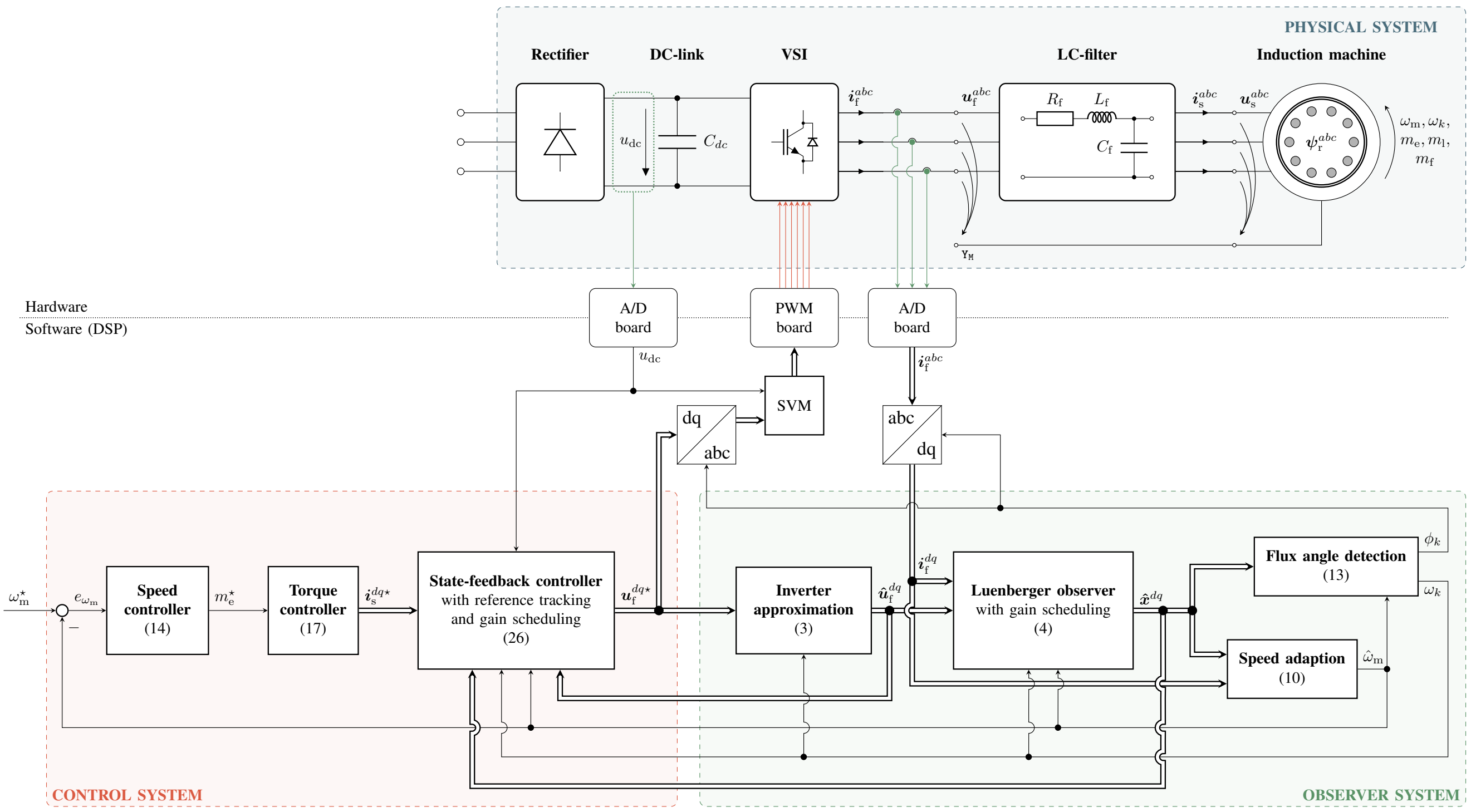

Fig. 1: Components of the electrical drive system comprising power electronics, LC-filter and induction machine.

### *B. Luenberger observer with gain-scheduling*

In the following a Luenberger observer is designed for the system described by (1). Since the VSI output voltages are generally not measured, the following assumption shall hold:

**Assumption 5** (Inverter output voltage)**.** *The inverter output voltage is known sufficiently well from approximation* (3)*, i.e.* $\hat{\boldsymbol{u}}_{\mathrm{f}}^{dq} \approx \boldsymbol{u}_{\mathrm{f}}^{dq}$ *holds.*

In view of Assumption 5, the observer dynamics are similar to (1) and can be stated as

$$\left.\begin{aligned} \tfrac{\mathrm{d}}{\mathrm{d}t}\hat{\boldsymbol{x}}^{dq} &= \boldsymbol{A}_{\mathrm{x}\to\mathrm{x}}(\hat{\omega}_{\mathrm{r}},\omega_k)\hat{\boldsymbol{x}}^{dq} + \boldsymbol{A}_{\mathrm{u}\to\mathrm{x}}\hat{\boldsymbol{u}}_{\mathrm{f}}^{dq} + \boldsymbol{\Lambda}_{\mathrm{x}}(\boldsymbol{y}_{\mathrm{x}}^{dq} - \hat{\boldsymbol{y}}_{\mathrm{x}}^{dq}) \\ \hat{\boldsymbol{y}}_{\mathrm{x}}^{dq} &= \boldsymbol{C}_{\mathrm{x}}\hat{\boldsymbol{x}}^{dq} (= \hat{\boldsymbol{i}}_{\mathrm{f}}^{dq}) \end{aligned}\right\} \quad (4)$$

with observed states $\hat{\boldsymbol{x}}^{dq} \in \mathbb{R}^8$, outputs $\hat{\boldsymbol{y}}_{\mathrm{x}}^{dq} \in \mathbb{R}^2$ and observer gain matrix $\boldsymbol{\Lambda}_{\mathrm{x}} \in \mathbb{R}^{8\times 2}$. Note that the observer has two inputs, namely the (approximated) inverter output voltage $\hat{\boldsymbol{u}}_{\mathrm{f}}^{dq} \approx \boldsymbol{u}_{\mathrm{f}}^{dq}$ and the measurement input $\boldsymbol{y}_{\mathrm{x}}^{dq}$. Since the angular velocity $\omega_{\mathrm{r}}$ is not measured, the system matrix $\boldsymbol{A}_{\mathrm{x}\to\mathrm{x}}(\hat{\omega}_{\mathrm{r}},\omega_k)$ further depends on the estimates $\hat{\omega}_{\mathrm{r}}$ and $\omega_k$.

The estimation error is defined as $\boldsymbol{e}_{\mathrm{x}}^{dq} := \boldsymbol{x}^{dq} - \hat{\boldsymbol{x}}^{dq} \in \mathbb{R}^8$ and the respective error dynamics—considering Assumption 5 and assuming perfect parameter knowledge, except for the mechanical speed—are given by

$$\tfrac{\mathrm{d}}{\mathrm{d}t}\boldsymbol{e}_{\mathrm{x}}^{dq} = \left[\boldsymbol{A}_{\mathrm{x}\to\mathrm{x}}(\omega_{\mathrm{r}},\omega_k) - \boldsymbol{\Lambda}_{\mathrm{x}}\boldsymbol{C}_{\mathrm{x}}\right]\boldsymbol{e}_{\mathrm{x}}^{dq} + \widetilde{\boldsymbol{A}}_{\mathrm{x}\to\mathrm{x}}\hat{\boldsymbol{x}}e_{\omega_{\mathrm{r}}} \quad (5)$$

with system error matrix

$$\widetilde{\boldsymbol{A}}_{\mathrm{x}\to\mathrm{x}} := \begin{bmatrix} \boldsymbol{0}_{2\times2} & \boldsymbol{0}_{2\times2} & \boldsymbol{0}_{2\times2} & \boldsymbol{0}_{2\times2} \\ \boldsymbol{0}_{2\times2} & \boldsymbol{0}_{2\times2} & \boldsymbol{0}_{2\times2} & \boldsymbol{0}_{2\times2} \\ \boldsymbol{0}_{2\times2} & \boldsymbol{0}_{2\times2} & \boldsymbol{0}_{2\times2} & -\frac{1}{L_m}\boldsymbol{J} \\ \boldsymbol{0}_{2\times2} & \boldsymbol{0}_{2\times2} & \boldsymbol{0}_{2\times2} & \boldsymbol{J} \end{bmatrix} \in \mathbb{R}^{8\times8}$$

and speed estimation error $e_{\omega_{\mathrm{r}}} := \omega_{\mathrm{r}} - \hat{\omega}_{\mathrm{r}} \in \mathbb{R}$.

Most publications dealing with speed-adaptive observers for induction machines focus on the proper selection of the feedback gains $\boldsymbol{\Lambda}_{\mathrm{x}}$, in order to achieve stable operation over a wide operation range (see e.g. [14], [15]). In this paper, a programmatic approach is chosen, which is based on the *linear quadratic regulator* (LQR) theory, in combination with the concept of gain-scheduling as to account for the time-varying parameters $\omega_k$ and $\hat{\omega}_{\mathrm{r}}$.

Since the eigenvalues of $\boldsymbol{A}_{\mathrm{x}\to\mathrm{x}} - \boldsymbol{\Lambda}_{\mathrm{x}}\boldsymbol{C}_{\mathrm{x}}$ are equal to the eigenvalues of its transpose $(\boldsymbol{A}_{\mathrm{x}\to\mathrm{x}} - \boldsymbol{\Lambda}_{\mathrm{x}}\boldsymbol{C}_{\mathrm{x}})^{\top}$ (dual system), finding the observer gain matrix $\boldsymbol{\Lambda}_{\mathrm{x}}$ can be reduced to a control problem, i.e. finding the optimal feedback gain matrix for the dual system

$$\tfrac{\mathrm{d}}{\mathrm{d}t}\breve{\boldsymbol{x}}_{\mathrm{x}}^{dq} = \boldsymbol{A}_{\mathrm{x}\to\mathrm{x}}(\hat{\omega}_{\mathrm{r}},\omega_k)^{\top}\breve{\boldsymbol{x}}_{\mathrm{x}}^{dq} + \boldsymbol{C}_{\mathrm{x}}{}^{\top}\breve{\boldsymbol{u}}^{dq}, \quad (6)$$

with dual state vector $\breve{\boldsymbol{x}}_{\mathrm{x}}^{dq} \in \mathbb{R}^8$, dual input vector $\breve{\boldsymbol{u}}^{dq} = -\boldsymbol{\Lambda}_{\mathrm{x}}^{\top}\breve{\boldsymbol{x}}_{\mathrm{x}}^{dq} \in \mathbb{R}^2$ and feedback gain matrix $\boldsymbol{\Lambda}_{\mathrm{x}}^{\top} \in \mathbb{R}^{2\times8}$ (see [16, Ch. 22.9]). The objective of the LQR approach is to minimize the cost function

$$J = \int_0^t \left[(\breve{\boldsymbol{x}}_{\mathrm{x}}^{dq})^{\top}\boldsymbol{Q}_{\Lambda}\breve{\boldsymbol{x}}_{\mathrm{x}}^{dq} + (\breve{\boldsymbol{u}}^{dq})\boldsymbol{R}_{\Lambda}\breve{\boldsymbol{u}}^{dq}\right]\mathrm{d}\tau, \quad (7)$$

by solving the algebraic Ricatti equation [17, Ch. 10.8]. The weighting matrices $\boldsymbol{Q}_{\Lambda} \in \mathbb{R}^{8\times8}$ and $\boldsymbol{R}_{\Lambda} \in \mathbb{R}^{2\times2}$, can be chosen in the following simple manner: (i) Use diagonal matrices for $\boldsymbol{Q}_{\Lambda}$ and $\boldsymbol{R}_{\Lambda}$, (ii) normalize the diagonal elements with respect to the *nominal* state or input variables (subscript ‘N’) and (iii) introduce a weighting factor $\alpha_{\Lambda} \in (0,1)$ to prioritize state weighting matrix $\boldsymbol{Q}_\Lambda$ or input weighting matrix $\boldsymbol{R}_\Lambda$. The following weighting matrices

$$\left.\begin{aligned} \boldsymbol{Q}_\Lambda &= \alpha_\Lambda \begin{bmatrix} \frac{1}{\|\boldsymbol{i}^{dq}_{\mathrm{f,nom}}\|^2}\boldsymbol{I}_2 & \boldsymbol{0}_{2\times 2} & \boldsymbol{0}_{2\times 2} & \boldsymbol{0}_{2\times 2} \\ \boldsymbol{0}_{2\times 2} & \frac{1}{\|\boldsymbol{u}^{dq}_{\mathrm{s,nom}}\|^2}\boldsymbol{I}_2 & \boldsymbol{0}_{2\times 2} & \boldsymbol{0}_{2\times 2} \\ \boldsymbol{0}_{2\times 2} & \boldsymbol{0}_{2\times 2} & \frac{1}{\|\boldsymbol{i}^{dq}_{\mathrm{s,nom}}\|^2}\boldsymbol{I}_2 & \boldsymbol{0}_{2\times 2} \\ \boldsymbol{0}_{2\times 2} & \boldsymbol{0}_{2\times 2} & \boldsymbol{0}_{2\times 2} & \frac{1}{\|\boldsymbol{\psi}^{dq}_{\mathrm{r,nom}}\|^2}\boldsymbol{I}_2 \end{bmatrix}, \\ \boldsymbol{R}_\Lambda &= (1-\alpha_\Lambda)\frac{1}{\|\boldsymbol{i}^{dq}_{\mathrm{f,nom}}\|^2}\boldsymbol{I}_2. \end{aligned}\right\} \quad (8)$$

are chosen. Note that the only tuning factor is $\alpha_\Lambda$, which makes the observer design a straight forward task.

Due to the fact that the system matrix depends on the parameters $\hat{\omega}_\mathrm{r}$ and $\omega_k$, the observer gains must be adapted online , i.e. $\boldsymbol{\Lambda}_\mathrm{x} = \boldsymbol{\Lambda}_\mathrm{x}(\hat{\omega}_\mathrm{r}, \omega_k)$ holds. This is achieved by offline calculation of the observer gains using the LQR method for several (but constant) values of $\hat{\omega}_\mathrm{r}$ and $\omega_k$ and storing the results in individual look-up tables (LUTs). During operation, the feedback gains are updated in each sampling instant using a 2D interpolation technique based on the new values of $\hat{\omega}_\mathrm{r}$ and $\omega_k$.

### C. Speed adaption

Typically, a PI-controller is used for the online adaption of the speed estimate $\hat{\omega}_\mathrm{r}$ (e.g. [6], [7], [18]). Input to the controller is the "*error torque*" $\boldsymbol{e}_{\boldsymbol{i}_\mathrm{s}}\boldsymbol{J}\hat{\boldsymbol{\psi}}^{dq}_\mathrm{r}$, resulting from the IM stator current estimation error $\boldsymbol{e}_{\boldsymbol{i}_\mathrm{s}} := \boldsymbol{i}^{dq}_\mathrm{s} - \hat{\boldsymbol{i}}^{dq}_\mathrm{s}$ and the rotor flux linkage estimate $\hat{\boldsymbol{\psi}}^{dq}_\mathrm{r}$. However, since only the filter currents $\boldsymbol{i}^{dq}_\mathrm{f}$ are available for feedback (see Assumption 4), a different input to the PI-controller is needed. Salomäki et al. [6] were the first to propose the use of a slightly different error torque

$$\epsilon = \boldsymbol{e}_{\boldsymbol{i}_\mathrm{f}}\boldsymbol{J}\hat{\boldsymbol{\psi}}^{dq}_\mathrm{r} \quad (9)$$

depending on the *filter* current estimation error $\boldsymbol{e}_{\boldsymbol{i}_\mathrm{f}} := \boldsymbol{i}^{dq}_\mathrm{f} - \hat{\boldsymbol{i}}^{dq}_\mathrm{f}$ and $\hat{\boldsymbol{\psi}}^{dq}_\mathrm{r}$, leading to the adaption law

$$\hat{\omega}_\mathrm{r} = K_{\mathrm{p},\hat{\omega}_\mathrm{r}}\epsilon + K_{\mathrm{i},\hat{\omega}_\mathrm{r}}\int_0^t \epsilon\,\mathrm{d}\tau, \quad (10)$$

with proportional gain $K_{\mathrm{p},\hat{\omega}_\mathrm{r}} \in \mathbb{R}$ and integral gain $K_{\mathrm{i},\hat{\omega}_\mathrm{r}} \in \mathbb{R}$.

**Remark 1.** *The gains $K_{\mathrm{p},\hat{\omega}_\mathrm{r}}$ and $K_{\mathrm{i},\hat{\omega}_\mathrm{r}}$ have significant impact on observer stability and must be chosen with care. It has been found that large values of $K_{\mathrm{i},\hat{\omega}_\mathrm{r}}$ lead to a faster speed adaption and a more robust zero-crossing capability, while, at the same time, oscillations in the state estimate occur if the value becomes too large. These oscillations can be reduced by increasing $K_{\mathrm{p},\hat{\omega}_\mathrm{r}}$, until satisfactory results are obtained.*

Steady-state (indicated by $\bar{\square}$) analysis of the state estimation error $\boldsymbol{e}^{dq}_\mathrm{x}$—i.e. by setting $\frac{\mathrm{d}}{\mathrm{d}t}\boldsymbol{e}^{dq}_\mathrm{x} = \boldsymbol{0}_8$ and solving (5) for $\boldsymbol{e}^{dq}_\mathrm{x}$—shows that the current error substitution is indeed feasible. The steady-state error torques are given by

$$\bar{\boldsymbol{e}}_{\boldsymbol{i}_\mathrm{f}}\boldsymbol{J}\hat{\boldsymbol{\psi}}^{dq}_\mathrm{r} = \gamma^\star_{\boldsymbol{i}_\mathrm{f}}(\omega_\mathrm{r}, \omega_k)\|\bar{\boldsymbol{\psi}}^{dq}_\mathrm{r}\|^2 e_{\omega_\mathrm{r}}, \quad (11)$$

$$\bar{\boldsymbol{e}}_{\boldsymbol{i}_\mathrm{s}}\boldsymbol{J}\hat{\boldsymbol{\psi}}^{dq}_\mathrm{r} = \gamma^\star_{\boldsymbol{i}_\mathrm{s}}(\omega_\mathrm{r}, \omega_k)\|\bar{\boldsymbol{\psi}}^{dq}_\mathrm{r}\|^2 e_{\omega_\mathrm{r}}, \quad (12)$$

with speed dependent "constants" $\gamma^\star_{\boldsymbol{i}_\mathrm{f}}(\omega_\mathrm{r}, \omega_k) \in \mathbb{R}$ and $\gamma^\star_{\boldsymbol{i}_\mathrm{s}}(\omega_\mathrm{r}, \omega_k) \in \mathbb{R}$. Calculating both constants numerically (e.g. with parameters in Tab. I) for different values of $\omega_\mathrm{r}$ and $\omega_k$, reveals that $\gamma^\star_{\boldsymbol{i}_\mathrm{f}} \approx \gamma^\star_{\boldsymbol{i}_\mathrm{s}}$ (see Fig. 2). Clearly, small deviations are observed, yet no switch of sign occurs and, as a consequence, using the filter current error for the speed adaption is a viable substitution.

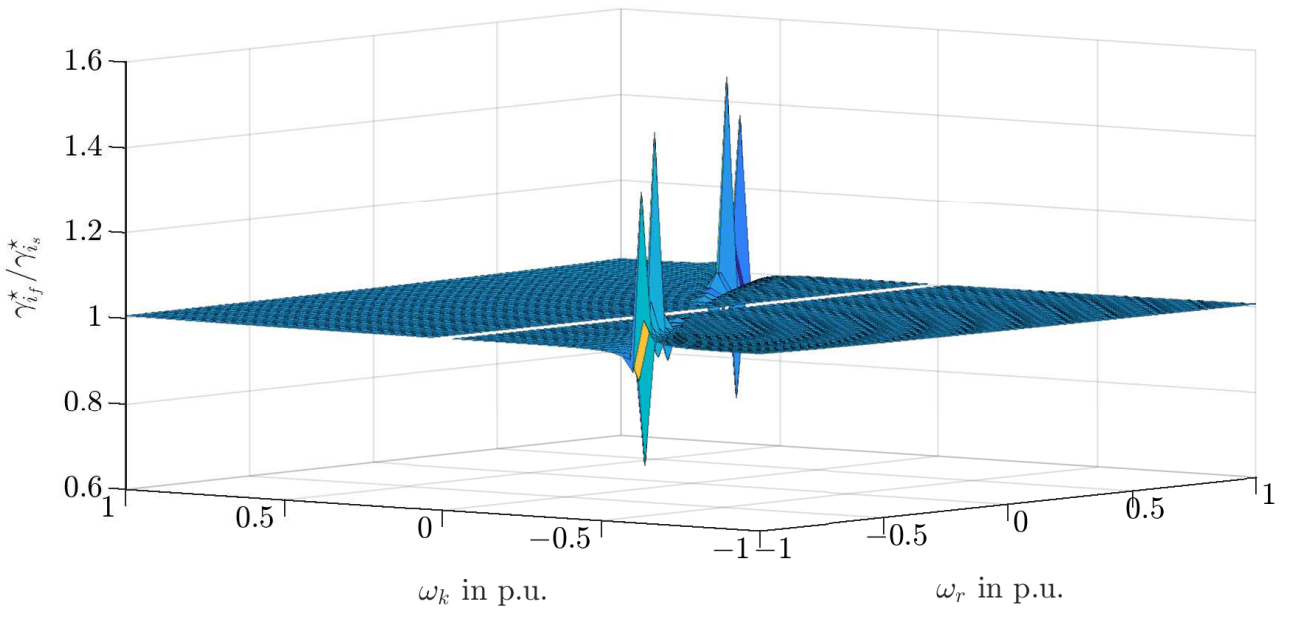

(a)

Fig. 2: Numerical calculation of the ratio $\gamma^\star_{\boldsymbol{i}_\mathrm{f}}/\gamma^\star_{\boldsymbol{i}_\mathrm{s}}$ of error torque constants $\gamma^\star_{\boldsymbol{i}_\mathrm{f}}$ and $\gamma^\star_{\boldsymbol{i}_\mathrm{s}}$ for different $\omega_\mathrm{r}$ and $\omega_k$ (excluding $\omega_\mathrm{r} = 0\,\mathrm{rad\,s^{-1}}$).

**Remark 2.** *It is only shown here that the stator torque error can be substituted by the filter current error, given the* standard *speed adaption law. However, the well-known stability problem in the* low *speed generating mode is not covered by the presented approach. Since the focus of this paper is on the control system rather than on the speed-adaption itself, the authors refer to the following publications for further details on that matter: [7], [14], [18], [19].*

### D. Flux angle detection

For rotor flux linkage orientation of the rotating $dq$-reference system, the speed $\omega_k$ of the synchronously rotating reference frame *and* the respective angle $\phi_k$ have to be defined. By assuming that the entire flux linkage is concentrated in the $d$-component, it follows that $\psi^q_\mathrm{r} = 0\,\mathrm{Wb}$ and $\frac{\mathrm{d}}{\mathrm{d}t}\psi^q_\mathrm{r} = 0\,\mathrm{V}$. Hence, the flux linkage dynamics of the $q$-component can be solved for $\omega_k$, yielding

$$\omega_k = \hat{\omega}_\mathrm{r} + \frac{L_m}{T_\mathrm{r}}\frac{\hat{i}^q_\mathrm{s}}{\hat{\psi}^d_\mathrm{r}} \quad \text{and} \quad \phi_k = \int_0^t \omega_k\,\mathrm{d}\tau. \quad (13)$$

### E. Stability of the observer

Stability proofs of speed-adaptive observers have extensively discussed in literature, since a first observer was used for induction machines by Kubota et al. [20] in the early 90s. Kubota tried to prove stability using a Lyapunov function yielding a simple integral adaption rule. However, the proof neglected an immeasurable flux term which weakens its validity. At the same time, other authors—e.g. Schauder [21] and Yang [22]—tried to prove stability using the concept of *hyperstability*. As pointed out by e.g. Suwankawin [23] in 2002, their proofs were wrong. In 2007, Sangwongwanich et al. were, as the first, able to provide a proof for *complete* stability under strict conditions on the observer gains [14]. Harnefors and Hinkkanen used a different approach to proof complete stability, by using a linearization approach [15]. To conclude: Proving complete stability is considered to be solved under strict conditions but not in general. Nevertheless, since the focus of this work is on the overall state-feedback control system and not on a rigor stability proof, the authors refer to earlier publications by e.g. Salomäki [6] and to the experimental and simulative validation as a proof of concept.

## IV. Control System

The control system is depicted in the lower left part of Fig. 1. A classical PI-speed controller is used to control the machine speed by passing a torque reference to the underlying feed-forward torque controller. The torque controller translates the torque set point into stator current setpoints. Finally, the state-feedback controller actively controls the stator currents in compliance with the given set point values, while the remaining system states are controlled indirectly. Instead of using and tuning several cascaded PI controllers (as e.g. in [6]), the proposed state-feedback controller tuning is straight forward. Employing the LQR tuning method ensures an easy design process and guarantees (overall) closed-loop stability.

### *A. Proportional-integral speed controller with anti-windup*

In view of the limited machine torque available, the considered PI-controller is implemented with anti-windup and is given by the following control law (see [13, Sec. 10.4.1])

$$\left.\begin{aligned} m_{\mathrm{e}}^{\star} &= K_{\mathrm{p},\omega} e_{\omega_{\mathrm{m}}} + K_{\mathrm{i},\omega}\xi_\omega, \\ \tfrac{\mathrm{d}}{\mathrm{d}t}\xi_\omega &= f(m_{\mathrm{e}}^{\star}) e_{\omega_{\mathrm{m}}}, \end{aligned}\right\} \tag{14}$$

with proportional $K_{\mathrm{p},\omega}$ and integral $K_{\mathrm{i},\omega}$ gains, integrator output $\xi_\omega$ and control error $e_{\omega_{\mathrm{m}}} := \omega_{\mathrm{m}}^{\star} - \omega_{\mathrm{m}}$. The anti-windup decision function (conditional integration)

$$f(m_{\mathrm{e}}^{\star}) := \begin{cases} 0, & \text{for } |m_{\mathrm{e}}^{\star}| > m_{\mathrm{e,max}} \\ 1, & \text{else.} \end{cases} \tag{15}$$

disables integration in (14) when the absolute value $|m_{\mathrm{e}}^{\star}|$ of the machine torque set point exceeds the maximally available IM torque $m_{\mathrm{e,max}}$ to suppress windup effects. Tuning of the gains $K_{\mathrm{p},\omega}$ and $K_{\mathrm{i},\omega}$ can be done by trial and error or by using analytical methods like the symmetrical optimum criterion.

### *B. Feed-forward torque controller*

The reference torque $m_{\mathrm{e}}^{\star}$ can be mapped to a pair of stator current set points $i_{\mathrm{s}}^{d^\star}$ and $i_{\mathrm{s}}^{q^\star}$. Typically, in the non-field weakening operation regime, a constant flux is used, i.e. the $d$-component of the stator currents is fixed to its nominal value, whereas the $q$-component is used to realize the given torque reference. The motor torque can be stated as

$$m_{\mathrm{e}} = \tfrac{3}{2} n_{\mathrm{p}} \tfrac{L_m}{L_{\mathrm{r}}} \psi_{\mathrm{r}}^{d} i_{\mathrm{s}}^{q}. \tag{16}$$

Hence, for constant $\psi_{\mathrm{r}}^{d\star}$ and given torque reference $m_{\mathrm{e}}^{\star}$, the $d-$ and $q$-current references become

$$i_{\mathrm{s}}^{d^\star} = \sin\varphi_N i_{\mathrm{s},N} \qquad \text{and} \qquad i_{\mathrm{s}}^{q^\star} = \frac{m_{\mathrm{e}}^{\star}}{\frac{3}{2} n_{\mathrm{p}} \frac{L_m}{L_r} \psi_{\mathrm{r}}^{d^\star}}, \tag{17}$$

where $\sin\varphi_N$ can be calculated from the power factor and $i_{\mathrm{s},N}$ is the rated current of the machine. Note that $i_{\mathrm{s}}^{q^\star}$ is only valid for non-zero values of $\psi_{\mathrm{r}}^{d\star}$.

### *C. State-feedback control of the drive system*

A state-feedback controller is designed which controls all system states simultaneously. Instead of using and tuning several and cascaded PI-controllers (as e.g. in [6]), the state-feedback controller tuning is simple and holistic. Employing the well-known LQR tuning method in combination with gain-scheduling ensures an easy design process and guarantees (overall) closed-loop stability. Control objective is reference tracking of the stator currents of the electric machine and suppression of oscillations in the remaining system states (due to the LC filter). Based on the *Separation Principle* (see e.g. [16, Ch. 18.4]), the controller is designed for system (1); assuming all states are available for feedback. Once the controller is derived, controller *and* observer are merged and implemented altogether.

*1) Augmented system:* The classical state-feedback controller does not allow for tracking of state reference values. Therefore, a new input $\boldsymbol{i}_{\mathrm{s}}^{dq\star} \in \mathbb{R}^2$ is defined, which represents the set point vector of the stator currents. Note that, since the system has two inputs only, merely two states can be controlled independently.

In order to ensure asymptotic set point tracking, i.e. $\lim_{t\to\infty} \ \boldsymbol{i}_{\mathrm{s}}^{dq} - \boldsymbol{i}_{\mathrm{s}}^{dq\star} = \boldsymbol{0}_2$, the system is extended by two additional states $\boldsymbol{\xi}_i^{dq} \in \mathbb{R}^2$ which represent the respective integrals over the state tracking errors. The dynamics of the new states are defined as

$$\tfrac{\mathrm{d}}{\mathrm{d}t}\boldsymbol{\xi}_i^{dq} = \boldsymbol{A}_{\mathrm{x}\to\mathrm{i}}\boldsymbol{x}^{dq} - \boldsymbol{i}_{\mathrm{s}}^{dq\star} = \boldsymbol{i}_{\mathrm{s}}^{dq} - \boldsymbol{i}_{\mathrm{s}}^{dq\star}, \tag{18}$$

where $\boldsymbol{A}_{\mathrm{x}\to\mathrm{i}} := [\boldsymbol{0}_{2\times4}, \boldsymbol{I}_2, \boldsymbol{0}_{2\times2}] \in \mathbb{R}^{2\times8}$ selects the stator currents $\boldsymbol{i}_{\mathrm{s}}^{dq}$ from the state vector $\boldsymbol{x}^{dq}$.

If, in addition, the inverter delay as approximated by (3) is considered, $\boldsymbol{u}_{\mathrm{f}}^{dq\star}$ (instead of $\boldsymbol{u}_{\mathrm{f}}^{dq}$) becomes the new control input and the overall augmented system can be stated as

$$\left.\begin{aligned} \tfrac{\mathrm{d}}{\mathrm{d}t}\bar{\boldsymbol{x}}^{dq} &= \bar{\boldsymbol{A}}(\hat{\omega}_{\mathrm{r}},\omega_k)\bar{\boldsymbol{x}}^{dq} + \bar{\boldsymbol{B}}\boldsymbol{u}_{\mathrm{f}}^{dq\star} + \bar{\boldsymbol{E}}\boldsymbol{i}_{\mathrm{s}}^{dq\star} \\ \boldsymbol{y}_c^{dq} &= \bar{\boldsymbol{C}}\bar{\boldsymbol{x}}^{dq} (= \boldsymbol{i}_{\mathrm{s}}^{dq}) \end{aligned}\right\} \tag{19}$$

where $\bar{\boldsymbol{x}}^{dq} := (\hat{\boldsymbol{u}}_{\mathrm{f}}^{dq^\top}, \boldsymbol{x}^{dq^\top}, \boldsymbol{\xi}_i^{dq^\top})^\top \in \mathbb{R}^{12}$ and $\boldsymbol{y}_c^{dq} := \boldsymbol{i}_{\mathrm{s}}^{dq} \in \mathbb{R}^2$ are the augmented state vector and control output, resp., and

$$\left.\begin{aligned} \bar{\boldsymbol{A}}(\hat{\omega}_{\mathrm{r}},\omega_k) &:= \left[\begin{array}{c|c|c} \boldsymbol{A}_{\mathrm{u}\to\mathrm{u}}(\omega_k) & \boldsymbol{0}_{2\times8} & \boldsymbol{0}_{2\times2} \\ \hline \boldsymbol{A}_{\mathrm{x}\to\mathrm{u}} & \boldsymbol{A}_{\mathrm{x}\to\mathrm{x}}(\hat{\omega}_{\mathrm{r}},\omega_k) & \boldsymbol{0}_{8\times2} \\ \hline \boldsymbol{0}_{2\times2} & \boldsymbol{A}_{\mathrm{x}\to\mathrm{i}} & \boldsymbol{0}_{2\times2} \end{array}\right] \\ \bar{\boldsymbol{B}} &:= \left[\boldsymbol{B}^\top \mid \boldsymbol{0}_{2\times8} \mid \boldsymbol{0}_{2\times2}\right]^\top, \\ \bar{\boldsymbol{E}} &:= \left[\boldsymbol{0}_{2\times2} \mid \boldsymbol{0}_{2\times8} \mid \boldsymbol{I}_2\right]^\top, \\ \bar{\boldsymbol{C}} &:= \left[\boldsymbol{0}_{2\times2} \mid \boldsymbol{A}_{\mathrm{x}\to\mathrm{i}} \mid \boldsymbol{0}_{2\times2}\right], \end{aligned}\right\} \tag{20}$$

are the augmented system matrix ($\bar{\boldsymbol{A}} \in \mathbb{R}^{12\times12}$), input matrix ($\bar{\boldsymbol{B}} \in \mathbb{R}^{12\times2}$), set point matrix ($\bar{\boldsymbol{E}} \in \mathbb{R}^{12\times2}$) and control output matrix ($\bar{\boldsymbol{C}} \in \mathbb{R}^{2\times12}$).

*2) State-feedback control law:* Since the augmented system (19) is fully controllable (can be shown, but omitted due to space limitations), the control law is

$$\boldsymbol{u}_{\mathrm{f}}^{dq\star} = -\bar{\boldsymbol{K}}\bar{\boldsymbol{x}}^{dq}, \tag{21}$$

with feedback gain matrix $\bar{\boldsymbol{K}} := [\boldsymbol{K}_\mathrm{u} \,|\, \boldsymbol{K}_\mathrm{x} \,|\, \boldsymbol{K}_\mathrm{i}] \in \mathbb{R}^{2\times 12}$, where $\boldsymbol{K}_\mathrm{u} \in \mathbb{R}^{2\times 2}$, $\boldsymbol{K}_\mathrm{x} \in \mathbb{R}^{2\times 8}$ and $\boldsymbol{K}_\mathrm{i} \in \mathbb{R}^{2\times 2}$. Similar to the observer gain selection, the feedback gain is calculated using the LQR approach with the following weighting matrices

$$\left.\begin{aligned}
\bar{\boldsymbol{Q}}_\mathrm{K} &= \alpha_K \begin{bmatrix} \mathbf{0}_{2\times 2} & \mathbf{0}_{2\times 2} & \mathbf{0}_{2\times 2} & \mathbf{0}_{2\times 2} & \mathbf{0}_{2\times 2} & \mathbf{0}_{2\times 2} \\ \mathbf{0}_{2\times 2} & \frac{\boldsymbol{I}_2}{\|\boldsymbol{i}_\mathrm{f,nom}^{dq}\|^2} & \mathbf{0}_{2\times 2} & \mathbf{0}_{2\times 2} & \mathbf{0}_{2\times 2} & \mathbf{0}_{2\times 2} \\ \mathbf{0}_{2\times 2} & \mathbf{0}_{2\times 2} & \frac{\boldsymbol{I}_2}{\|\boldsymbol{u}_\mathrm{s,nom}^{dq}\|^2} & \mathbf{0}_{2\times 2} & \mathbf{0}_{2\times 2} & \mathbf{0}_{2\times 2} \\ \mathbf{0}_{2\times 2} & \mathbf{0}_{2\times 2} & \mathbf{0}_{2\times 2} & \frac{\boldsymbol{I}_2}{\|\boldsymbol{i}_\mathrm{s,nom}^{dq}\|^2} & \mathbf{0}_{2\times 2} & \mathbf{0}_{2\times 2} \\ \mathbf{0}_{2\times 2} & \mathbf{0}_{2\times 2} & \mathbf{0}_{2\times 2} & \mathbf{0}_{2\times 2} & \frac{\boldsymbol{I}_2}{\|\boldsymbol{\psi}_\mathrm{r,nom}^{dq}\|^2} & \mathbf{0}_{2\times 2} \\ \mathbf{0}_{2\times 2} & \mathbf{0}_{2\times 2} & \mathbf{0}_{2\times 2} & \mathbf{0}_{2\times 2} & \mathbf{0}_{2\times 2} & \beta_K \boldsymbol{I}_2 \end{bmatrix}, \\
\boldsymbol{R}_\mathrm{K} &= (1-\alpha_K)\frac{1}{\|\boldsymbol{u}_\mathrm{f,nom}^{dq}\|^2}\boldsymbol{I}_2.
\end{aligned}\right\} \quad (22)$$

The factor $\beta_K$ in $\bar{\boldsymbol{Q}}_\mathrm{K}$ constitutes an additional tuning factor weighting the integral-action.

*3) Controller gain-scheduling:* Similarly to the observer gain scheduling described in Sec. III, the controller gains are updated in each control cycle. Likewise, the LQR algorithm has to be repeated offline for several combinations of $\omega_\mathrm{r}$ and $\omega_k$, yielding 2D LUTs for each entry of $\bar{\boldsymbol{K}}$ and, hence, $\bar{\boldsymbol{K}} = \bar{\boldsymbol{K}}(\hat{\omega}_\mathrm{r}, \omega_k)$ holds.

*4) Prefilter:* So far, the integral-action introduced in the previous section was the only means to achieve set point tracking. However, in order to achieve faster tracking, it is advisable to further add a prefilter to the control law. Let $\boldsymbol{V} \in \mathbb{R}^{2\times 2}$ be the prefilter matrix, then the modified control law including the prefilter is given as follows

$$\boldsymbol{u}_\mathrm{f}^{dq\star} = -\bar{\boldsymbol{K}}\bar{\boldsymbol{x}}^{dq} + \boldsymbol{V}\boldsymbol{i}_\mathrm{s}^{dq\star}. \quad (23)$$

where the prefilter matrix is defined as (see e.g. [24, Ch. 7.5])

$$\boldsymbol{V}(\hat{\omega}_\mathrm{r},\omega_k) := -\left(\left[\mathbf{0}_{2\times 2} \mid \boldsymbol{A}_{\mathrm{x}\to\mathrm{i}}\right]\left[\begin{array}{c|c} \boldsymbol{A}_{\mathrm{u}\to\mathrm{u}}(\omega_k) - \boldsymbol{B}\boldsymbol{K}_\mathrm{u} & -\boldsymbol{B}\boldsymbol{K}_\mathrm{x} \\ \hline \boldsymbol{A}_{\mathrm{x}\to\mathrm{u}} & \boldsymbol{A}_{\mathrm{x}\to\mathrm{x}}(\hat{\omega}_\mathrm{r},\omega_k) \end{array}\right]^{-1}\left[\begin{array}{c} \boldsymbol{B} \\ \hline \mathbf{0}_{8\times 2} \end{array}\right]\right)^{-1}. \quad (24)$$

A detailed derivation of the prefilter matrix is omitted here due to space limitations. Note, though, that $\boldsymbol{V}(\hat{\omega}_\mathrm{r}, \omega_k)$ depends on $\hat{\omega}_\mathrm{r}$ and $\omega_k$, which requires its recalculation in each control step or the use of an additional LUT.

*5) Output saturation:* Since the output voltage of the VSI is constrained, the controller output must be limited, too. Therefore, the magnitude of the reference voltage is limited by

$$\boldsymbol{u}_\mathrm{f,sat}^{dq\star} = \begin{cases} \boldsymbol{u}_\mathrm{f}^{dq\star} & \text{, for } \|\boldsymbol{u}_\mathrm{f}^{dq\star}\| \le u_\mathrm{f,max}(u_\mathrm{dc}) \\ \boldsymbol{u}_\mathrm{f}^{dq\star} \cdot \frac{u_\mathrm{f,max}(u_\mathrm{dc})}{\|\boldsymbol{u}_\mathrm{f}^{dq\star}\|} & \text{, else,} \end{cases} \quad (25)$$

where the maximum voltage $u_\mathrm{f,max}(u_\mathrm{dc}) \in \mathbb{R}$ depends on the dc-link voltage and the employed modulation scheme (e.g. for space vector modulation, $u_\mathrm{f,max} = u_\mathrm{dc}/\sqrt{3}$ holds). The saturated output $\boldsymbol{u}_\mathrm{f,sat}^{dq\star} \in \mathbb{R}^2$ is passed on to the modulator.

### D. Implementation of the overall system

Having derived the observer and controller systems independently, they can finally be merged into a single system, yielding the overall system dynamics

$$\begin{aligned}\frac{\mathrm{d}}{\mathrm{d}t}\hat{\bar{\boldsymbol{x}}}^{dq} &= \big(\bar{\boldsymbol{A}}(\hat{\omega}_\mathrm{r},\omega_k) - \bar{\boldsymbol{\Lambda}}(\hat{\omega}_\mathrm{r},\omega_k)\bar{\boldsymbol{C}}\big)\hat{\bar{\boldsymbol{x}}}^{dq} + \bar{\boldsymbol{B}}\boldsymbol{u}_\mathrm{f,sat}^{dq\star} \\ &\quad + \bar{\boldsymbol{E}}\boldsymbol{i}_\mathrm{s}^{dq\star} + \bar{\boldsymbol{\Lambda}}(\hat{\omega}_\mathrm{r},\omega_k)\boldsymbol{i}_\mathrm{f}^{dq},\end{aligned} \quad (26)$$

TABLE I: Parameters of the test setup.

| | Parameter | Variable | Value | Unit |
|---|---|---|---|---|
| **VSI** | DC-link voltage | $u_\mathrm{dc}$ | 580 | V |
| | Switching frequency | $f_\mathrm{sw}$ | 4000 | Hz |
| **Filter** | Rated current (amplitude) | $\hat{i}_\mathrm{f,N}$ | 22 | A |
| | Inductance | $L_\mathrm{f}$ | $4.5\times 10^{-3}$ | H |
| | Capacitance | $C_\mathrm{f}$ | $30\times 10^{-6}$ | F |
| | Resistance | $R_\mathrm{f}$ | 0.1 | Ω |
| **Induction machine** | Rated speed (nameplate) | $\omega_\mathrm{m,N}$ | 298.4 | $\mathrm{rad\,s^{-1}}$ |
| | Rated torque | $m_\mathrm{m,N}$ | 10.05 | N m |
| | Rated voltage (amplitude) | $u_\mathrm{s,N}$ | 327 | V |
| | Rated current (amplitude) | $i_\mathrm{s,N}$ | 8.1 | A |
| | Rated power factor | $\cos(\varphi_\mathrm{N})$ | 0.93 | |
| | Rated flux (amplitude) | $\psi_\mathrm{r,N}$ | 1.2 | Wb |
| | Number of pole pairs | $n_\mathrm{p}$ | 1 | |
| | Stator resistance | $R_\mathrm{s}$ | 1.85 | Ω |
| | Rotor resistance | $R_\mathrm{r}$ | 1.55 | Ω |
| | Main inductance | $L_m$ | $340\times 10^{-3}$ | H |
| | Stator leakage inductance | $L_{\mathrm{s}\sigma}$ | $16.5\times 10^{-3}$ | H |
| | Rotor leakage inductance | $L_{\mathrm{r}\sigma}$ | $16.5\times 10^{-3}$ | H |
| **Control system** | P-gain (speed estimator) | $K_{\mathrm{p},\hat{\omega}_\mathrm{r}}$ | -1 | |
| | I-gain (speed estimator) | $K_{\mathrm{i},\hat{\omega}_\mathrm{r}}$ | -120 | |
| | P-gain (speed control) | $K_{\mathrm{p},\omega}$ | 1.2 | |
| | I-gain (speed control) | $K_{\mathrm{i},\omega}$ | 1 | |
| | 1. weighting factor (obs.) | $\alpha_\Lambda$ | $1.5\times 10^{-5}$ | |
| | 1. weighting factor (contr.) | $\alpha_K$ | 0.5 | |
| | 2. weighting factor (contr.) | $\beta_K$ | 1 | |

with state vector $\hat{\bar{\boldsymbol{x}}}^{dq} := (\hat{\boldsymbol{u}}_\mathrm{f}^{dq}, \hat{\boldsymbol{x}}^{dq}, \boldsymbol{\xi}_i^{dq})^\top \in \mathbb{R}^{12}$, extended observer matrix $\bar{\boldsymbol{\Lambda}} := [\mathbf{0}_{2\times 2}, (\boldsymbol{\Lambda}_\mathrm{x})^\top, \mathbf{0}_{2\times 2}]^\top \in \mathbb{R}^{12\times 2}$ and control law $\boldsymbol{u}_\mathrm{f}^{dq\star}$ as defined in . The overall system (26) can be used for the implementation. Note that the observer matrix $\boldsymbol{\Lambda}_\mathrm{x}(\hat{\omega}_\mathrm{r}, \omega_k)$ and control matrix $\bar{\boldsymbol{K}}(\hat{\omega}_\mathrm{r}, \omega_k)$ are updated using look-up tables in each control step, according to the actual speed estimate $\hat{\omega}_\mathrm{r}$ and the synchronous speed $\omega_k$.

## V. Experimental & simulative validation

In this section, simulative and experimental validation of the proposed control scheme are shown for a system described by the parameters given in Table I. Since the controller is supposed to run on a digital signal processing (DSP) unit, the derived control system has to be discretized. The implementation itself is done in MATLAB and Simulink R2017a for both, simulation and experiment. The simulation environment, as well as the experimental setup are briefly described. Finally, the results are discussed and evaluated.

### A. Discretization

In the preceding sections, the continuous-time state-space system (26) was derived. Let

$$\frac{\mathrm{d}}{\mathrm{d}t}\boldsymbol{x}(t) = \boldsymbol{A}_t\boldsymbol{x}(t) + \boldsymbol{B}_t\boldsymbol{u}(t) \quad (27)$$

be a generic continuous-time linear system with state vector $\boldsymbol{x} \in \mathbb{R}^n$, system matrix $\boldsymbol{A}_t \in \mathbb{R}^{n\times n}$, input $\boldsymbol{u} \in \mathbb{R}^m$ and input matrix $\boldsymbol{B}_t \in \mathbb{R}^{n\times m}$. It can be shown—under the assumption, that the input does not change during one sampling period—that its discrete equivalent is given by (see [25, Sec. 2.6.3])

$$\boldsymbol{x}[k+1] = \boldsymbol{A}_k\boldsymbol{x}[k] + \boldsymbol{B}_k\boldsymbol{u}[k], \quad (28)$$

with system and input matrices

$$\left.\begin{aligned} \boldsymbol{A}_k &:= \exp\left(\boldsymbol{A}_t t_{\mathrm{sw}}\right) = \boldsymbol{I}_n + \boldsymbol{S}_\infty \boldsymbol{A}_t \\ \boldsymbol{B}_k &:= \big(\exp\left(\boldsymbol{A}_t t_{\mathrm{sw}}\right) - \boldsymbol{I}_n\big)\boldsymbol{A}_t^{-1}\boldsymbol{B}_t = \boldsymbol{S}_\infty \boldsymbol{B}_t. \end{aligned}\right\} \tag{29}$$

and, for $N = \infty$, $\boldsymbol{S}_N := t_{\mathrm{sw}} \sum_{\nu}^{N} \boldsymbol{A}_t^{\nu} \frac{t_{\mathrm{sw}}^{\nu}}{(\nu+1)!}$. Typically, the series expansion is neglected after $\nu = 0$ (i.e. $\boldsymbol{S}_0$), yielding the commonly used explicit (forward) Euler method. However, as experiments showed, this first order discretization is insufficient for the given setup. Chosing $N = 1$ (second order) gave good results for a reasonable sampling time $t_{\mathrm{sw}} = 250\,\mu\mathrm{s}$. Hence, the following discretized model

$$\begin{aligned} \hat{\bar{\boldsymbol{x}}}^{dq}[k+1] &= \boldsymbol{I}_{12} + \boldsymbol{S}_1(\hat{\omega}_{\mathrm{r}}, \omega_k)\big(\bar{\boldsymbol{A}}(\hat{\omega}_{\mathrm{r}}, \omega_k) - \bar{\boldsymbol{\Lambda}}(\hat{\omega}_{\mathrm{r}}, \omega_k)\bar{\boldsymbol{C}}\big)\hat{\bar{\boldsymbol{x}}}^{dq}[k] \\ &\quad + \boldsymbol{S}_1(\hat{\omega}_{\mathrm{r}}, \omega_k)\bar{\boldsymbol{B}}\boldsymbol{u}_{\mathrm{f,sat}}^{dq\star}[k] + \boldsymbol{S}_1(\hat{\omega}_{\mathrm{r}}, \omega_k)\bar{\boldsymbol{\Lambda}}(\hat{\omega}_{\mathrm{r}}, \omega_k)\boldsymbol{i}_{\mathrm{f}}^{dq}[k] \end{aligned} \tag{30}$$

is used for the implementation. Note that the LQR algorithm has to be applied on the discretized system. For example, using *MATLAB*, the function `dlqr(...)` of the *Control System Toolbox* has to be used, instead of `lqr(...)`.

### *B. Simulation*

The control system is implemented as a discrete block in Simulink, triggered at the center of each PWM period, just as in the experimental setup. The two-level VSI is supplied by a constant DC-link voltage, while the switching signals are generated using space-vector modulation (SVM). The LC filter and the (linear) induction machine are simulated based on model (1), while continuous-time integrators are used to solve the first-order differential equations. Moreover, a simplified mechanical model with viscous friction and arbitrary load torque is used. The fixed-step solver `ode3` runs with a sampling time of $100\,\mathrm{ns}$.

### *C. Experimental setup*

The testbench (see Fig. 3) comprises a $3\,\mathrm{kW}$ induction and load machine, both equipped with position encoders, a torque sensor, a custom-built LC filter, two 2-level VSIs and the dSPACE real-time system.The modular dSPACE system runs on a DS1007 processing unit, with a DS5101 module for the PWM generation, a DS2004 A/D module, and a DS3002 encoder board. Note that, unlike stated in the beginning of this chapter, stator currents, voltages and rotor speed are measured here. However, this data is only used for evaluation; it is *not* fed back to the control system.

### *D. Results & Discussion*

Simulations and experiments have been conducted for two different scenarios: (i) constant load torque and varying speed (see Fig. 4) and (ii) constant speed and varying load torque (see Fig. 5). The results simulation and experimental results are presented in the left and right columns, respectively; in the first row, the control scenario is depicted, with the reference speed $\omega_{\mathrm{m}}^{\star}$ and load torques $m_{\mathrm{l}}^{\star}$, and the respective measured $(\omega_{\mathrm{m}}, m_{\mathrm{m}})$ and estimated values $(\hat{\omega}_{\mathrm{m}})$. The second to fifth rows show the measured and estimated system states, e.g. filter currents $\boldsymbol{i}_{\mathrm{f}}^{dq}$, stator voltages $\boldsymbol{u}_{\mathrm{s}}^{dq}$, stator currents $\boldsymbol{i}_{\mathrm{s}}^{dq}$ and rotor flux linkages $\boldsymbol{\psi}_{\mathrm{r}}^{dq}$ (from top to bottom). Since the flux linkages could not be measured, only the estimated values are shown in the respective plot.

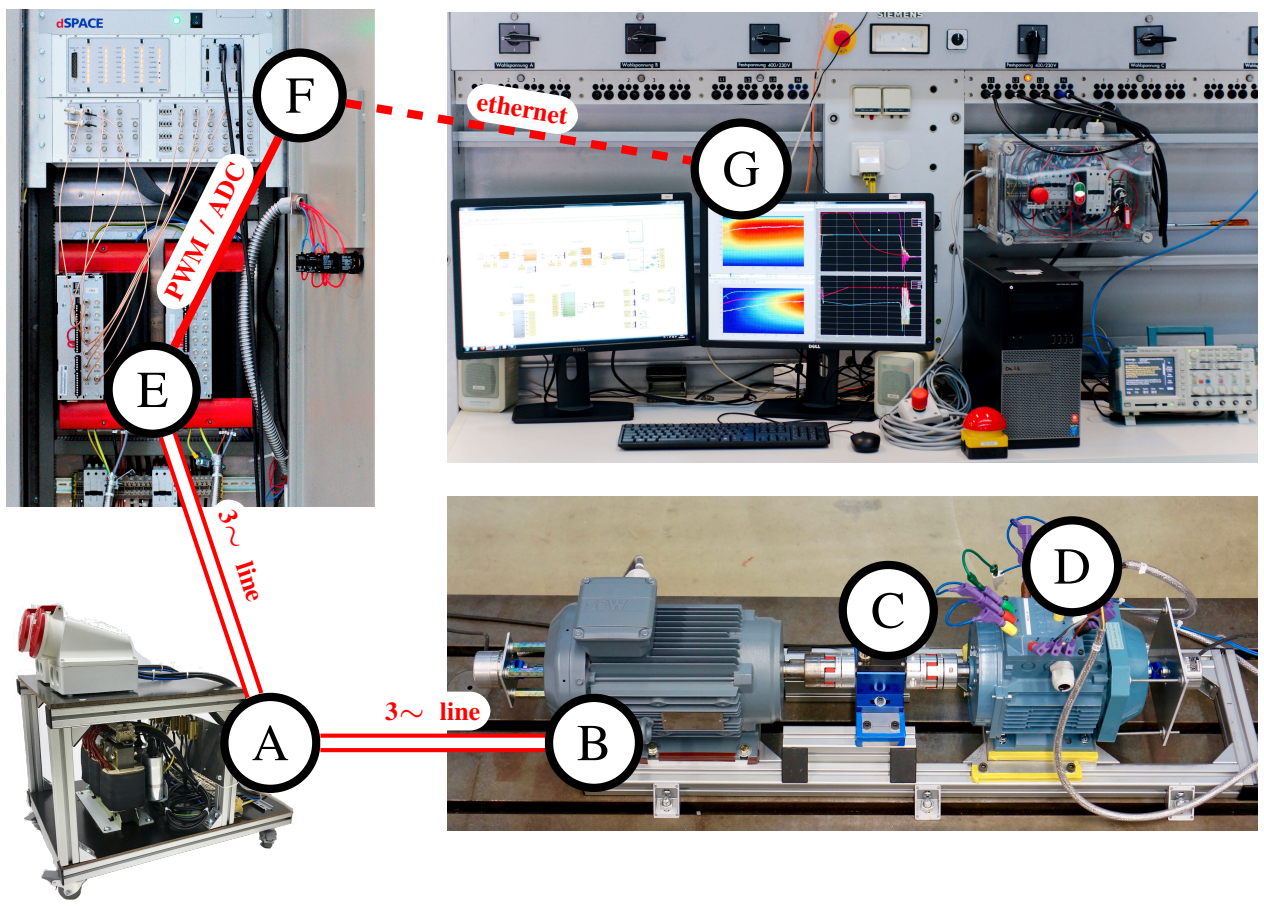

Fig. 3: Testbench: (A) LC filter, (B) IM, (C) torque sensor, (D) load machine, (E) VSI, (F) real-time system, (G) host PC.

*1) Scenario (i): Constant load torque, varying speed:* In the first scenario, the speed $\omega_{\mathrm{m}}$ varies between rated positive and negative speed, including zero-crossings at $t \approx 12.5\,\mathrm{s}$ and $t \approx 21.5\,\mathrm{s}$, while the load torque jumps from zero to rated torque at $t = 3\,\mathrm{s}$. Prior to this jump, the system is initialized, i.e. the flux $\psi_{\mathrm{r}}^{d}$ is controlled to its rated value and the speed to its initial reference value. Generally, the speed estimator performs well, except for two short periods observed at the speed zero-crossings. Here, the observer looses track due to the well-known instability region in the low speed generating regime (see Remark 2). As a consequence, the system states are not tracked properly here, too. However, since the integral gain $K_{\mathrm{i},\hat{\omega}_{\mathrm{r}}}$ of the speed-adaption was chosen sufficiently large, the observer is able to return to normal operation after a brief period of time (see Remark 1). As for the torque measurement, it can be observed that the experiment shows slightly smoother results than the simulation which might be explained by internal low-pass filtering of the torque transducer. Moreover, the measured motor torque $m_{\mathrm{m}}$ deviates slightly from the load torque reference $m_{\mathrm{l}}^{\star}$, which can be explained by friction acting either constructively or destructively (viscous friction was incorporated in the simulation model), depending on the mode of operation: In motoring mode, the (speed controlled) IM has to deliver the sum of load and frictional torque, whereas in generating mode it is the difference between load and frictional torque.

Looking at the system states, the simulation and experimental data match very well in most parts. In both cases, the system states are decently tracked. Only the zero-crossing areas and sections of very high torque and currents, where the controller saturates, can be considered critical. It is worth mentioning, though, that the instability of the speed estimation is less severe in the experimental results. A possible explanation is the internal low-pass filtering of the current sensors (filter time constant $20\,\mu\mathrm{s}$), which smoothes the current measurement and thereby possibly smoothes the speed estimation, too.

Nonetheless, the experimental results are generally more noisy which is explained by the varying DC-link voltage, being upper-bound by the chopper resistor in generating mode and lower-bound by the rectifier in motoring mode. In turn, for the simulations, a constant DC-link voltage has been assumed.

Another striking observation is that the filter and stator $d$-currents, although stabilized, are not estimated properly, which is caused by the nonlinear flux saturation in real machines (not considered in the linear IM model). This hypothesis is supported by the simulation results, where saturation has not been modeled and the estimation error tends to zero.

*2) Scenario (ii): Constant speed, varying load:* In the second scenario, the flux and speed initialization is omitted and the speed is kept constant at its rated value, while the load torque is varied between its positive and negative rated value. The results show similar characteristics as observed in scenario (i), yet overall a very good match between simulation and experimental results is observed and the observer achieves asymptotic tracking over the whole load spectrum.

## VI. Conclusion

A speed-sensorless state-feedback control system for an induction machine with LC filter has been derived. Its validity was verified in simulations and experiments, which showed a good overall match and decent controller and estimator performance. The main advantage of the presented approach is its easy implementation, including a tuning approach, which relies on well-known methods (e.g. LQR) and described heuristics. Moreover, practical advice on the implementation (i.e. weighting factor selection & discretization) has been given. Future work comprises stability improvements in the zero-speed range and considerations about a nonlinear flux model as to improve the angle, current and flux estimation.

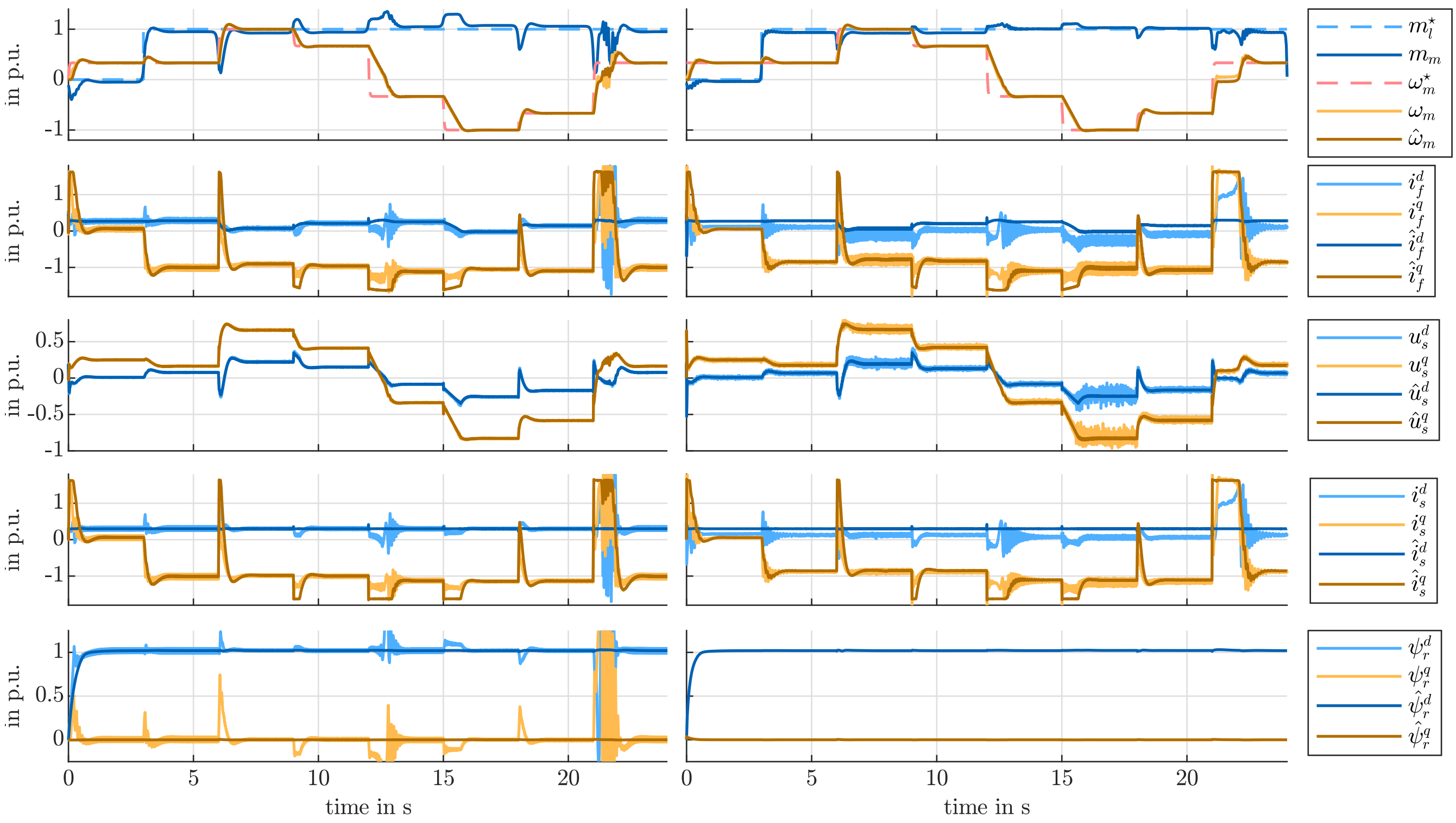

(a) (b)

Fig. 4: (a) Simulation and (b) experimental results for scenario (i): constant torque and varying speed.

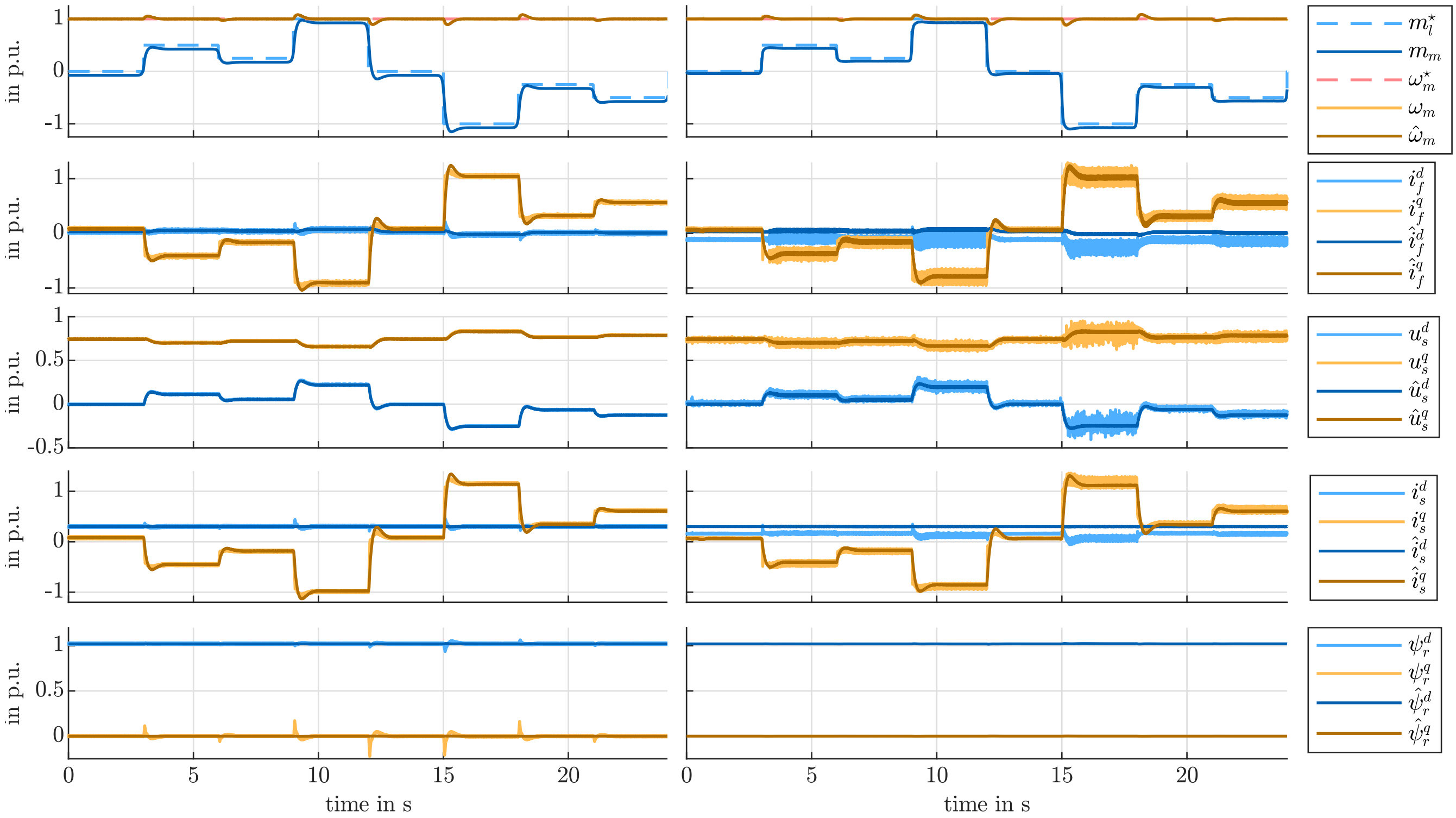

(a) (b)

Fig. 5: (a) Simulation and (b) experimental results for scenario (ii): varying torque and constant speed.